\documentclass[a4paper,11pt]{elsarticle}
\usepackage[brazil]{babel}
\usepackage{amsmath,amsthm,amsfonts,amssymb}
\usepackage[mathcal]{eucal}
\usepackage{mathrsfs}
\usepackage{graphicx}
\usepackage{dcolumn}
\usepackage{mathtools}
\usepackage{latexsym}
\usepackage{epsfig}
\usepackage{bm}
\usepackage[all]{xy}
\usepackage[T1]{fontenc}
\newcommand{\ie}{\begin{equation}}
\newcommand{\fe}{\end{equation}}
\usepackage{url}

\usepackage[figurename=Fig.]{caption}
\newcommand{\sech}{\mathrm{sech} \,}
\usepackage{amsmath}
\usepackage{amsthm}
\usepackage{amsfonts}
\usepackage{amssymb}
\usepackage{slashed}
\usepackage{xfrac}
\usepackage{graphicx}
\usepackage{dsfont}
\usepackage[mathcal]{eucal}
\usepackage{indentfirst}
\usepackage{graphicx}
\usepackage{parskip}
\usepackage{bigints}
\usepackage{nicefrac}
\newcommand\fverb{\setbox\fverbbox=\hbox\bgroup\verb}
\newcommand\fverbdo{\egroup\medskip\noindent%
            \fbox{\unhbox\fverbbox}\ }
\newcommand\fverbit{\egroup\item[\fbox{\unhbox\fverbbox}]}
\newbox\fverbbox

\def\text#1{\mbox{#1}}

\journal{Physics Letters B}

\begin{document}
\begin{frontmatter}
\title{Fermions in a warped resolved conifold.}
\author{D. M. Dantas, J. E. G. Silva, C. A. S. Almeida}
\address{Departamento de F\'{i}sica - Universidade Federal do Cear\'{a} \\ C.P. 6030, 60455-760
Fortaleza-Cear\' {a}-Brazil}

\begin{abstract}
We investigated the localization of the spinorial field in a braneworld built as a warped product between a 3-brane and a 2-cycle of the resolved conifold. This scenario provides a geometric flow that controls the singularity at the origin and changes the properties of the fermion in this background. Furthermore, due the cylindrical symmetry according to the 3-brane and a smoothed warp factor, this geometry can be regarded as a near brane correction of the string-like branes. This geometry allows a normalizable and well-defined massless mode whose decay and value on the brane depend on the resolution parameter. For the Kaluza-Klein modes, resolution parameter also controls the height of the barrier of the volcano potential.
\end{abstract}


\end{frontmatter}
\section{Introduction}

The Randall-Sundrum model changed the way we understand the universe by allowing the spacetime to have infinite extra dimensions \cite{Randall:1999ee,Randall:1999vf}. In spite of the localization of the gravity on the 3-brane, the gauge and fermion fields are not trapped in this model \cite{Kehagias:2000au}. One way to overcome this issue is to extend the RS model to higher dimensions \cite{Olasagasti:2000gx}.

In six dimensions, a static space time with an infinite extra dimension and cylindrical symmetry is the so-called string-like model
\cite{Olasagasti:2000gx,Liu:2007gk, Oda:2000zc,Gherghetta:2000qi,Cohen:1999ia,Gregory:1999gv,Giovannini:2001hh, Tinyakov:2001jt, Ponton:2000gi}.
This model have the advantage of localize the massless mode of both fermions \cite{Liu:2007gk} and gauge \cite{Oda:2000zc} fields on the brane coupled with only the gravity. Furthermore, the correction to the gravitational potential is less than in RS model \cite{Gherghetta:2000qi}. However, due the conical behavior near the brane, the string-like model has the problem of find non-zero induced field equations on the brane \cite{Bostock:2003cv}.

Another important property of the string-like model is the relationship between physics and geometry. Indeed, the geometry of
the transverse manifold, as its deficit angle, is related to the mass-tension of the string-brane
\cite{Olasagasti:2000gx,Gherghetta:2000qi,Giovannini:2001hh}. This effect motivated us to study how the fields on these models are affected by
some geometrical flow in the extra dimensions.

We performed this task choosing as a parameter dependent transverse manifold a 2-cycle of the so-called resolved conifold. This smooth six
dimensional space whose parameter $a$ controls the singularity on the tip of the cone is a special internal Calabi-Yau space of string-theory
\cite{Candelas:1989js, Greene:1995hu,p, Pando Zayas:2000sq,Cvetic:2000mh, Minasian:1999tt, Klebanov:2007us, VazquezPoritz:2001zt}. Thus, it is
possible continuously to pass from a smooth to a singular manifold varying the parameter $a$. This geometrical resolution flow
is also used in an extension of the AdS/CFT correspondence \cite{Pando Zayas:2000sq,Klebanov:2007us,Klebanov:1999tb,Klebanov:2000hb,Klebanov:2000nc}.

The study of the behavior of the fields on braneworlds with a resolved transverse conifold was addressed before in the literature. For the gravitational field, in a 10-dimensional space-time, the massless mode is located around the origin and the KK spectrum has an exponential decay \cite{VazquezPoritz:2001zt}. In a
six-dimensional set-up, we have shown that the scalar field has massless and massive modes trapped to the brane \cite{Silva:2011yk}. Moreover, the resolution flow changes the
properties of the volcano potential for the KK modes, as the width of the well and the height of the barrier \cite{Silva:2011yk}.

In this article we have used a different warp factor, firstly studied in \cite{Silva:2012yj}, and that possess a $Z_{2}$ symmetry. This warp function satisfies the required regularity conditions, what renders this geometry as a smooth extension of the string-like scenario. This geometry represents a positive tension brane embedded in a spacetime with negative cosmological constant \cite{Silva:2011yk}. Furthermore, for tiny values of
$a$ the components of the stress-energy tensor satisfy the weak and strong energy condition what extends the thin string-like model \cite{Gherghetta:2000qi,Tinyakov:2001jt}. On the other hand, for $a\neq 0$, the 3-brane can be regarded as a brane embedded in a 4-brane with a compact extra dimension whose radius is the resolution parameter. This enable us to realize the RS$1$ model
as a limit of the six dimensional non-compact scenario.

Once studied the geometry we turned our attention to the behavior of a massless spinorial field minimally coupled in this scenario. For the massless mode, it turned out that this mode
is normalizable provided there is a background gauge vector field, as done in \cite{Liu:2007gk}. Moreover, the new warp factor smooth out this mode at the origin
while the resolution parameter controls the value of the gauge field on the brane.

Another improvement obtained is related to the conical behavior of the string-like models. Indeed, the conical geometry yields a divergence of the zero mode on the brane. On the other hand, if we consider a thin string-brane, taking into account only the exterior geometry, the metric does not satisfies the regularity conditions \cite{Gherghetta:2000qi,Tinyakov:2001jt}. Also, it is possible to achieve a well-defined zero mode for others 6D conical geometries, but with compact
transverse space \cite{WILLIAMS:2012AU}. The resolution parameter prevents this singular effect by smoothing out the cone at the origin.

For the KK modes, there is an attractive potential for only the left-handed fermion \cite{Liu:2007gk}. As for the scalar field, the depth of the well and the height of the barrier of the usual volcano potential depend on the resolution parameter \cite{Silva:2011yk}. Nevertheless, despite the lack of a potential well at the origin for the right-handed fermion, there is a potential well besides the brane.

This work is organized as follows. In section \ref{Bulk geometry} we built the warped product between a $3$-brane and the $2$-cycle of the conifold and studied the geometric properties of this scenario. In section \ref{Fermionic zero modes} we have studied the properties of the massless and KK spectrum of the fermionic field. Some conclusions and perspectives are outlined in section \ref{Conclusions and perspectives}.


\section{Bulk geometry}
\label{Bulk geometry}

Consider a six dimensional warped bulk
$\mathcal{M}_{6}$ of form $\mathcal{M}_{6}=\mathcal{M}_{4}\times \mathcal{M}_{2}$, where $\mathcal{M}_{4}$ is a 3-brane embedded in $\mathcal{M}_{6}$ and $\mathcal{M}_{2}$ is a two-dimensional transverse space.

The action for this model is defined as
\begin{equation}
\label{action}
  S_{g} =\int_{\mathcal{M}_{6}}{\left(\frac{1}{2\kappa_{6}}R-\Lambda +\mathcal{L}_{m}\right)\sqrt{-g}d^{6}x},
\end{equation}
where $\kappa_{6}=\frac{8\pi}{M_{6}^{4}}$, $M_{6}^{4}$ is the six-dimensional bulk Planck mass and $\mathcal{L}_{m}$ is the source matter Lagrangian.

\begin{figure}[htb] 
       \begin{minipage}[b]{0.48 \linewidth}
           \fbox{\includegraphics[width=\linewidth]{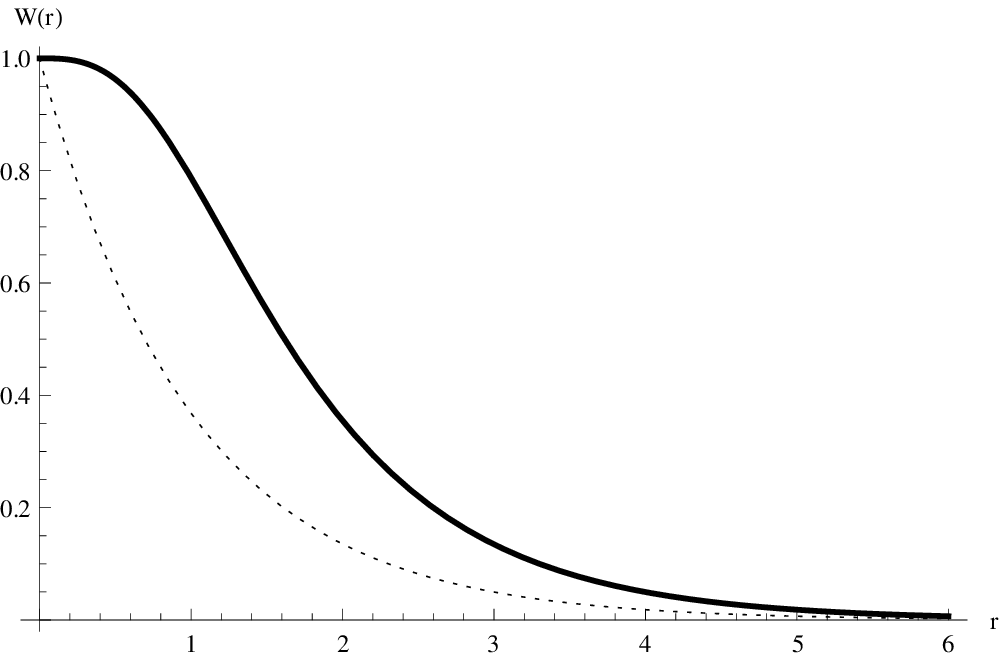}}\\
           \caption{Warp function for $c=1$ (thick line). The thin string warp factor (dotted line) is defined only for the exterior of the string.}
           \label{warp factor}
       \end{minipage}\hfill
       \begin{minipage}[b]{0.48 \linewidth}
           \fbox{\includegraphics[width=\linewidth]{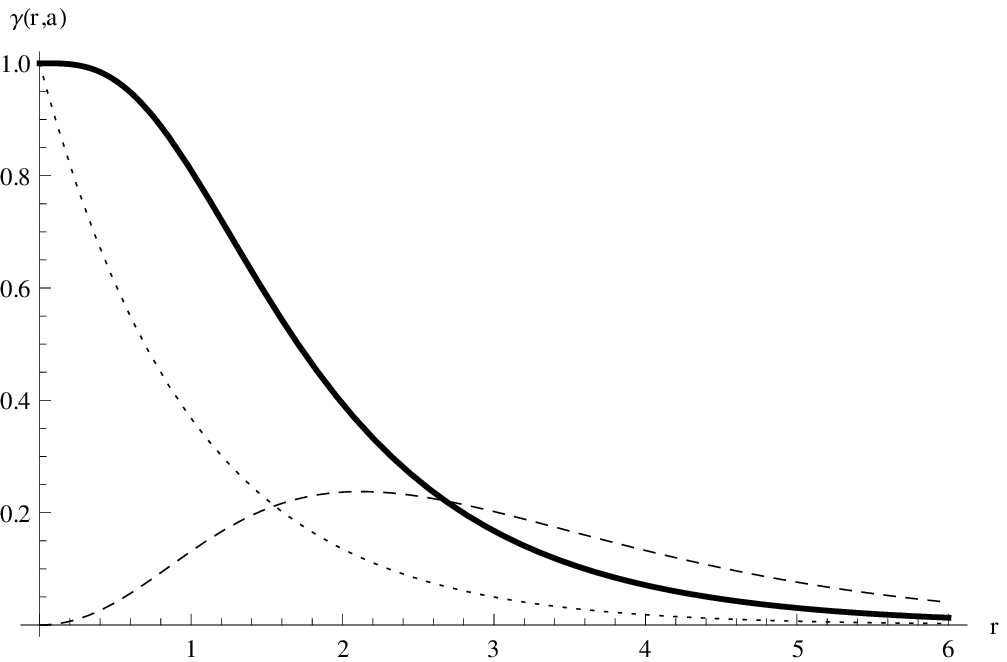}}\\
           \caption{Angular metric component for $c=1$. For $a=0$ (dashed line) there is a conical singularity and the thin string-like geometry is denoted by the dotted line.}
           \label{angularcomponent}
       \end{minipage}
   \end{figure}

Consider a static and axisymmetric warped metric between a flat 3-brane $\mathcal{M}_{4}$ and the transverse manifold $\mathcal{M}_{2}$ given by
\cite{Gherghetta:2000qi,Giovannini:2001hh,Tinyakov:2001jt,Silva:2012yj}
\begin{eqnarray}
\label{metricansatz}
ds^{2}_{6} & =  & W(r,c)\eta_{\mu\nu}dx^{\mu}dx^{\nu}+dr^{2} + \gamma(r,c,a) d\theta^{2},
\end{eqnarray}
where $W\in C^{\infty}$ is the so called warp factor. For the thin string-like models, the metric is given by \cite{Liu:2007gk,Oda:2000zc,Gherghetta:2000qi,Gregory:1999gv,Giovannini:2001hh,Tinyakov:2001jt,Ponton:2000gi}
\begin{eqnarray}
\label{thinstringlike}
W(r)= e^{-cr} & , & \gamma(r)= R_{0}^{2}e^{-cr},
\end{eqnarray}
where $c^{2}=-\frac{2K_{6}}{5}\Lambda$. The system in eq. (\ref{thinstringlike}) describes the exterior geometry of the defect. It can be understood as a warped product between a 3-brane and
a disc of radius $R_{0}$. Furthermore, the metric components in eq. (\ref{thinstringlike})
do not satisfy the regularity conditions at the origin, namely,
\begin{eqnarray}
 W(0,c)=1 & , & W'(0,c)= 0,
\end{eqnarray}
where, the prime $(')$ stands for the derivative $\frac{d}{dr}$. In order to overcome this problem, in this work, we shall use a smoothed warp factor \cite{Silva:2012yj,Costa:2013}
\begin{equation}
\label{warpfunction}
W(r,c)=e^{-(cr - \tanh{cr})}.
\end{equation}
The addition of the term $\tanh{c r}$ smoothes the warp factor near the origin, as shown in the fig. (\ref{warp factor}). Therefore, we can realize this warp function as a near brane correction to the thin string-like models \cite{Olasagasti:2000gx,Liu:2007gk,Oda:2000zc,Gherghetta:2000qi}.

Moreover, instead of use the disc, we have chosen a 2-section of the resolved conifold as the transverse manifold \cite{Candelas:1989js,Pando Zayas:2000sq,Cvetic:2000mh,VazquezPoritz:2001zt,Silva:2011yk}
\begin{eqnarray}
\label{transversespacemetric}
ds^{2}_{2} & = & \left(\frac{u^{2}+6a^{2}}{u^{2}+9a^{2}}\right)du^{2}+ \frac{1}{6}(u^{2}+6a^{2})d\theta^{2}.
\end{eqnarray}

Asymptotically, the resolved conifold has a conical shape. Near the origin the constant $a$, called the resolution parameter, controls the divergence of the conifold. This resolution flow provides a way to study the effects of a conical singularity has on the fields.

\begin{figure}[htb] 
       \begin{minipage}[b]{0.48 \linewidth}
           \fbox{\includegraphics[width=\linewidth]{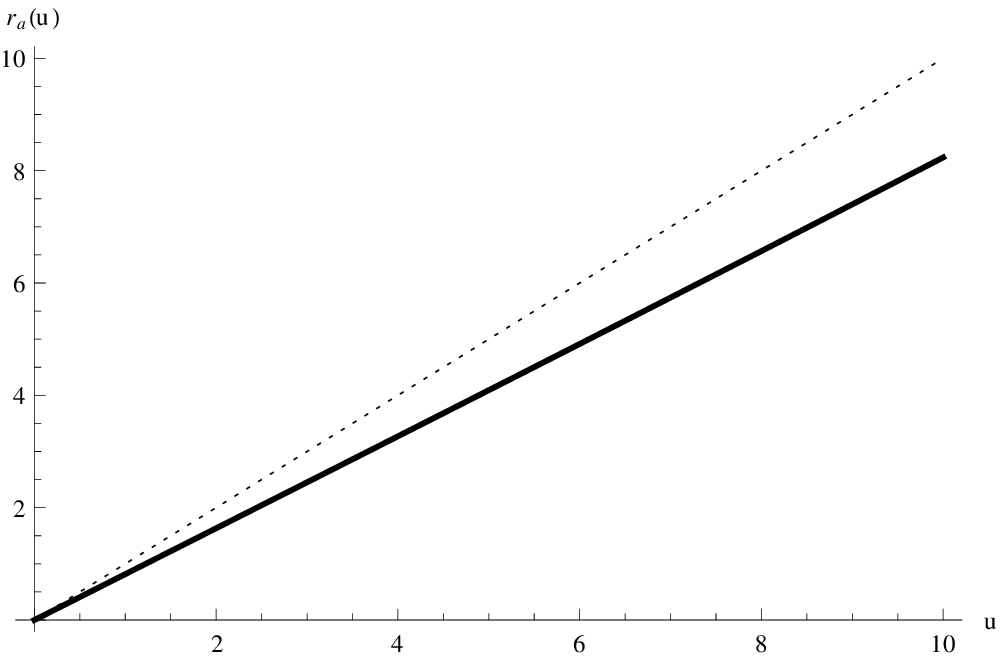}}\\
           \caption{Change of radial coordinate for $a=10$ (thick line) and for $a=0$ (dotted line).}
           \label{variablechange}
       \end{minipage}\hfill
       \begin{minipage}[b]{0.48 \linewidth}
           \fbox{\includegraphics[width=\linewidth]{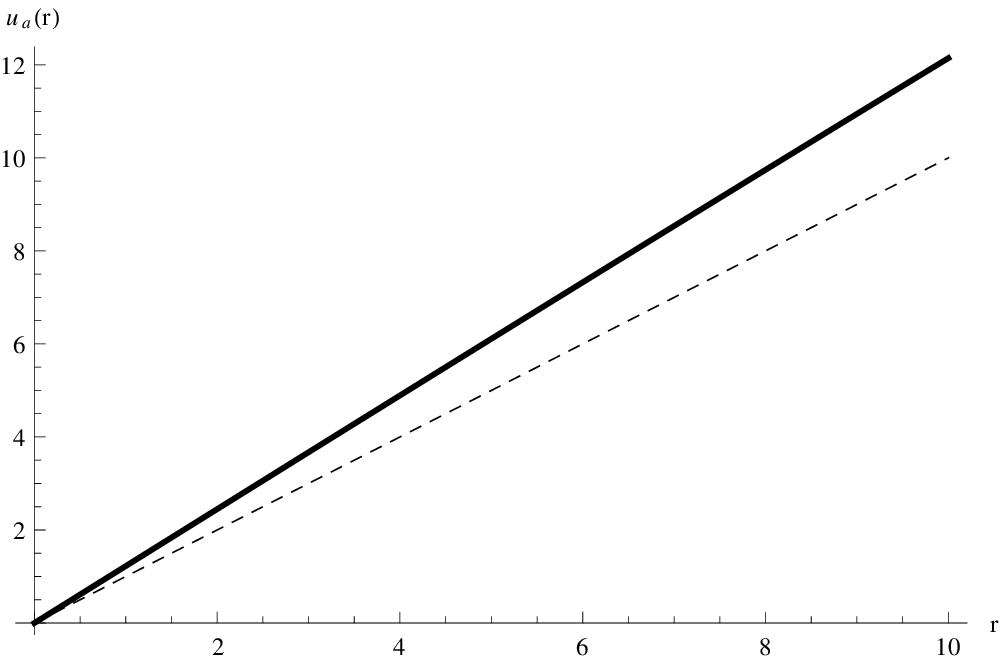}}\\
           \caption{Inverse change of variable for $a=10$ (thick line) and for $a=0$ (dashed line).}
           \label{changeofvariableinverse}
       \end{minipage}
\end{figure}

The coordinates $u$ and $r$ are related by
\begin{displaymath}
\label{diff}
 r_{a}(u) = \left\{
 \begin{array}{lr}
  u & ,a = 0\\
  -i\sqrt{6}a E\left(\text{arcsinh}\left(\frac{i}{3a}u\right),\frac{3}{2}\right) & , a\neq 0,
     \end{array}
  \right.
\end{displaymath}
whose behavior is sketched in the fig. (\ref{variablechange}).

For the angular metric component, $\gamma:[0,\infty) \rightarrow [0,\infty)$, we have modified the string-like ansatz using as metric
\cite{Gherghetta:2000qi,Giovannini:2001hh,Silva:2012yj, Costa:2013},
\begin{eqnarray}
\label{angularmetric}
\gamma(r,c,a)= W(r,c)\beta(r,a)=e^{-(cr - \tanh{cr})}\left(\frac{u(r,a)^{2}+6a^{2}}{6}\right).
\end{eqnarray}
The angular component (\ref{angularmetric}) provides an resolved conical behavior to the transverse manifold. At the origin, the angular component satisfies $\gamma(0,c,a)=a^{2}$. Then, the geometrical flow of the resolved conifold yields a dimensional reduction $\mathcal{M}_{6}\rightarrow \mathcal{M}_{5}$ at the origin. The string-like dimensional reduction $\mathcal{M}_{6}\rightarrow \mathcal{M}_{4}$ is achieved only for $a=0$. Therefore, the resolution flow connects the string-like models
(for $a=0$) and the RS1 model \cite{Randall:1999ee} for $a\neq 0$.

As shown in \cite{Costa:2013}, this scenario has a smooth scalar curvature and it converges asymptotically to an $AdS_{6}$ manifold. The metric ansatz $(5)$ and $(7)$ satisfy the Einstein equation for a source whose stress-energy components satisfy the weak and dominant energy
conditions \cite{Costa:2013}. A detailed analysis of Einstein equations, the string tensions and the relationship between the bulk and brane mass scales
can also be found in \cite{Costa:2013}.

\section{Fermion Localization}
\label{Fermionic zero modes}

In this section we shall study the effects that the resolution flow has on a Dirac fermion
in this scenario. Among the advantages are the existence of a well-defined zero-mode on the brane and a parametrization of the Schr\"{o}dinger potential.

Consider the action for a minimally coupled spin $\frac{1}{2}$ spinor \cite{Oda:2000zc, Liu:2007gk, Parameswaran:2007,Neronov:2001, Merab:2007, Aguilar:2006}, namely
\begin{eqnarray}
\label{action-fermions-6D}
 S_6=\int{\sqrt{-g}\bar{\Psi} i\Gamma^{M} D_M \Psi}d^6x,
\end{eqnarray}
where $\Gamma^{M}=\xi^{M}_{\bar{M}}\Gamma^{\bar{M}}$ are the curved Dirac matrices defined from the flat Dirac matrices $\Gamma^{\bar{M}}$ through the vielbein $\xi^{M}_{\bar{M}}$ and $D_M$ is the gauge covariant derivative given by \cite{Liu:2007gk, Parameswaran:2007}
\begin{eqnarray}\label{Dcov}
D_M=\partial_M +\frac{1}{4}\omega^{\bar{M} \bar{N}}_{M} \Gamma_{\bar{M}}\Gamma_{\bar{N}} - iqA_M,
\end{eqnarray}
where $A_{M}$ is a background gauge vector field.

In this geometry, the massless Dirac equation satisfies the equation
\begin{eqnarray}
\label{Dirac-6D-2}
\Gamma^M D_M\Psi= \Big[W^{-\frac{1}{2}}\Gamma^{\bar{\mu}}\Big(\partial_\mu-iqA_{\mu}(x)\Big)+\Gamma^{\bar{r}}\Big(\partial_r + \Big[\frac{W'}{W}+ \frac{\big( \beta W\big)'}{4\beta W}\Big]-iqA_r(r)\Big)+\nonumber \\+\big(\beta W\big)^{-\frac{1}{2}}\Gamma^{\bar{\theta}}\Big(\partial_{\theta}-iqA_{\theta}(r)\Big)\Big]=0.
\end{eqnarray}

Following the usual approach, we shall use the spinor representation \cite{Liu:2007gk, Oda:2000zc, Parameswaran:2007, Neronov:2001,Merab:2007, Aguilar:2006}
\begin{equation}
\label{rowspinor}
\Psi(x,r,\theta)=
\begin{pmatrix}
\psi_{4}\\
0\\
\end{pmatrix}_{8\times 1}
\end{equation}

\begin{eqnarray}
\label{Matrices-6D-4D}
\Gamma^{\bar{\mu}}=
\begin{pmatrix}
0 &\gamma^{\bar{\mu}}\\
\gamma^{\bar{\mu}} &0
\end{pmatrix}_{8\times 8},\quad
\Gamma^{\bar{r}}=
\begin{pmatrix}
0 &\gamma^{5}\\
\gamma^{5} &0
\end{pmatrix},\quad
\Gamma^{\bar{\theta}}=
\begin{pmatrix}
0 &-i\\
i &0\\
\end{pmatrix},
\end{eqnarray}
and $\gamma^5$ is such that $\gamma^{5}\psi_{R,L}=\pm \psi_{R,L}$.

Further, let us perform a Kaluza-Klein decomposition on $\psi_{4}$ in the form
\begin{equation}
\label{spinorkkdecomposition}
\psi_4(x,r,\theta)=\sum\limits_{l}[\psi_{R_l}(x)\alpha_{R_l}(r)+\psi_{L_l}(x)\alpha_{L_l}(r)]e^{il\theta}.
\end{equation}

Using equations (\ref{rowspinor}), (\ref{Matrices-6D-4D}), and (\ref{spinorkkdecomposition}), the Dirac equation (\ref{Dirac-6D-2}) turns to be
\begin{eqnarray}\label{Dirac-6D-3}
\Gamma^M D_M\Psi(x,r,\theta)= \sum\limits_{l}e^{il\theta}\Big[W^{-\frac{1}{2}}m\psi_{L_l,R_l}
\pm \Big(\partial_r + \frac{W'}{W}+ \frac{\big( \beta W\big)'}{4\beta W}-iqA_r(r)+\\ \nonumber
+(\beta W)^{-\frac{1}{2}}\big(qA_{\theta}(r)-l\big)\Big)\psi_{R_l,L_l}(x)\Big]\alpha_{R_l,L_l}(r)=0.
\end{eqnarray}

In this work, we will be concerned with the solutions for $l=0$, the so-called s-waves. In this case, eq. (\ref{Dirac-6D-3}) yields
\begin{eqnarray}\label{Dirac-6D-4}
\Big(\partial_r + \Big[\frac{W'}{W}+ \frac{\big( \beta W\big)'}{4\beta W}- iqA_r(r)\pm\frac{q}{\sqrt{\beta W}}A_{\theta}(r)\Big]\Big)\alpha_{R,L}(r)=\mp \frac{m}{\sqrt{W}} \alpha_{L,R}(r)
\end{eqnarray}

\subsection{Zeromode}
\label{zeromode}

Now let us study the solution of eq. (\ref{Dirac-6D-4}) for $m=0$, the so-called massless mode. Due to the difference of the expressions for $a=0$ and
$a\neq 0$, we shall split the analysis in two steps.

\subsubsection{Conical behavior $a=0$}

Using eq. (\ref{warpfunction}) and eq. (\ref{angularmetric}), the massless mode for $a=0$ (singular cone) is given by
\begin{eqnarray}
\label{Loc-zero-mode}
\alpha_{R,L}(r)=C_0 \frac{1}{\sqrt{r}}\exp{\Big[\frac{5}{4}[cr-\tanh(cr)]-q\int^{r}[iA_r(r') \pm \frac{ \sqrt{6}}{r'}e^{\frac{1}{2}[cr'-\tanh(cr')]}A_{\theta}(r')]dr'\Big]}.
\end{eqnarray}

The eq. (\ref{Loc-zero-mode}) is similar to the string-like one except for the factor $\frac{1}{\sqrt{r}}$ that prevents us to define an induced fermion on the brane \cite{Oda:2000zc,Liu:2007gk}. This result shows us that a non-singular cone is essential for the spinor be well-defined on this scenario. The problem of induced fields and field equation on string-like branes due the conical behavior near the origin is well-known for the scalar and gravitational field \cite{Silva:2011yk,Silva:2012yj}.

In spite of this, the zero-mode (\ref{Loc-zero-mode}) can leads to an effective four-dimensional action depending on the
form of the vector gauge field. In fact,
\begin{eqnarray}
\label{action-fermions-6D-2}
S_{6_{eff}}=\int_{x^M}\sqrt{-g}\bar{\Psi} i\Gamma^{M} D_M \Psi d^{M}x =2\pi\int_x d^{\mu}x\bar{\psi}i\Gamma^{\bar{\mu}}\partial_{\mu}\psi\int_0^{\infty}dr W^2\beta^{\frac{1}{2}}\lvert \alpha_{R,L}(r)\rvert^2.
\end{eqnarray}
Then, we need to analyze the integral
\begin{eqnarray}
\label{normal1}
I_{R,L}(r) & = & \int_0^{\infty}dr W^2\beta^{\frac{1}{2}}\lvert \alpha_{R,L}(r)\rvert^2\nonumber\\
                    & = & C_0^2\int_{0}^{\infty}dr{\big(e^{\frac{1}{2}[cr-\tanh(cr)]}\big)e^{\mp 2\sqrt{6}q\bigintsss^{r}{dr' A_{\theta}(r')\big[\frac{1}{r'}e^{\frac{1}{2}[cr'-\tanh(cr')]}\big]}}} .
\end{eqnarray}

It is worthwhile to say that the integral (\ref{normal1}) depends only on the angular gauge vector component $A_{\theta}$. Thus, let us
choose a general ansatz for the right-handed spinor in the form
\begin{eqnarray}
\label{atheta1}
A_{{\theta}_R}(r)=\Big(\frac{r e^{-\frac{1}{2}[cr-\tanh(cr)]}}{4 \sqrt{6} }\Big)\lambda \tanh^2 b(r-r_0),
\end{eqnarray}
where $\lambda$ and $b$ are free parameters. With the expression (\ref{atheta1}) the integral (\ref{normal1}) turns to be
\begin{eqnarray}
\label{normal2}
I_{R}=C^2\int_{0}^{\infty}dr{e^{\frac{1}{2}[(c-\lambda)r+\frac{\lambda}{b}\tanh b(r-r_0)-\tanh(cr)]}}.
\end{eqnarray}

\begin{figure}[htb] 
       \begin{minipage}[t]{0.48 \linewidth}
                  \fbox{\includegraphics[width=\linewidth]{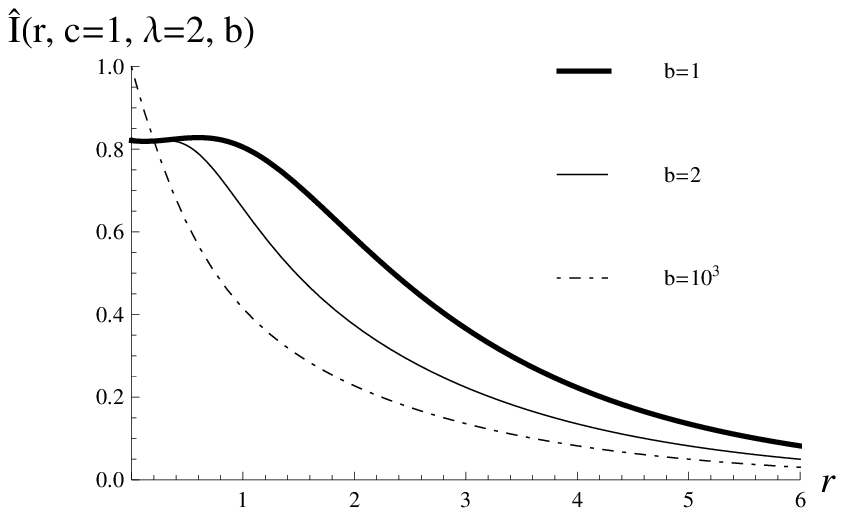}}\\
           \caption{$\hat{I}$ for $a=0$ and $r_0=0.1$.}
           \label{normal-fig}
       \end{minipage}\hfill
       \begin{minipage}[t]{0.48 \linewidth}
 \fbox{\includegraphics[width=\linewidth]{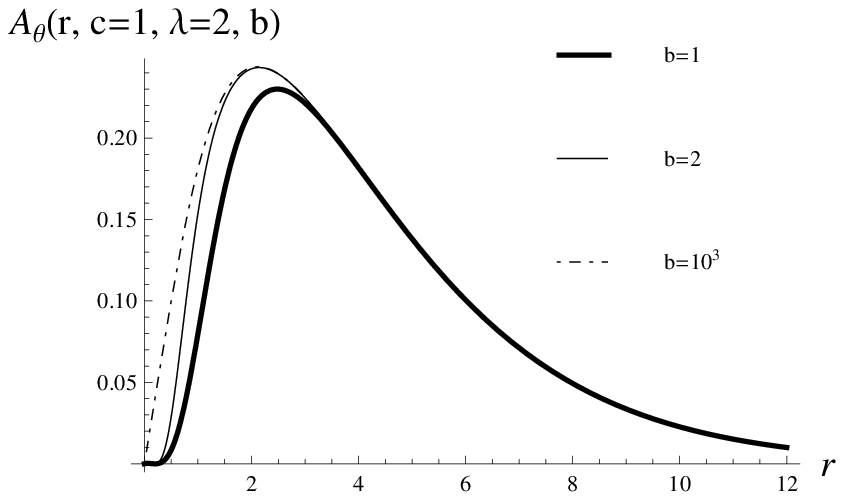}}\\
           \caption{Angular gauge component for $a=0$.}
           \label{atheta-fig}
       \end{minipage}
\end{figure}

In order to $I_{R}$ converges, we impose that $\lambda > c$  and for a smooth $I_{R}$ we further suppose that $\lambda\geq b>c$. Taking $c=1$, we plotted the graphics for the integrand of $I_{R}$, $\hat{I}$, and $A_{\theta}$
in Figures (\ref{normal-fig}) and (\ref{atheta-fig}). It is worthwhile to say that the background bulk gauge field $A_{\theta}$ has the same behavior of an Abelian vortex in six dimensions \cite{Giovannini:2001hh}.

For the massless left-handed fermion be localized it turns out that we need to shift the sign of $A_{\theta}$ in (\ref{atheta1}).
Thus, even though both modes yield an effective action, for a given gauge field $A$ there is only one localized mode.


\subsubsection{Resolved conifold $a\neq0$}

For $a\neq0$, the solution of eq. (\ref{Dirac-6D-4}) is given by
\begin{eqnarray}\label{Loc-zero-modeb}
\alpha_{R,L}(r)=\frac{C_a}{ \sqrt[4]{ v_{a}^2-1}} \exp {\Big[\frac{5}{4}[cr-\tanh(cr)]-q\int[iA_r(r) \pm \frac{ e^{\frac{1}{2}[cr-\tanh(cr)]}}{a\sqrt{1-v_a^2}}A_{\theta}(r)]dr\Big]},
\end{eqnarray}
where $C_a$ is a constant and $v_{a}= E\left(\text{arcsinh}\left(\frac{i}{\sqrt{6}a}r\right),\frac{2}{3}\right)$.

It is worthwhile to say that the zero-mode for $a\neq0$ is well-defined at the origin what solves the brane induced field problem.
The resolution parameter controls the value of this massless mode at the brane and it represents the radius of the compact dimension as well.

The form of $\alpha_{R,L}(r)$ in (\ref{Loc-zero-modeb}) yields the expression for $I_{R,L}$
\begin{eqnarray}\label{normal1b}
I_{R,L}(r)=C_a^2\int_{0}^{\infty}{dr (e^{\frac{1}{2}[cr-\tanh(cr)]}\big)e^{\mp 2q\bigintsss{dr \frac{e^{\frac{1}{2}[cr-\tanh(cr)]}}{a\sqrt{1-v_a^2}}A_{\theta}(r)}}}.
\end{eqnarray}

For the angular gauge component, we have chosen the following ansatz
\begin{eqnarray}\label{atheta1b}
A_{{\theta}_R}(r)=\Big(a\frac{\sqrt{1- v_a^2}e^{-\frac{1}{2}[cr-\tanh(cr)]}}{4  q}\Big)\lambda\tanh^2 b (u_a(r)-r_0),
\end{eqnarray}
where the constant $r_{0}$ yields a non-vanishing gauge field at the origin. Hence, $I_{R}$ assumes the form
\begin{eqnarray}\label{normal2b}
I_{R}=C^2\int_{0}^{\infty}dr{e^{\frac{1}{2}[(c-\lambda)r-\lambda\int{\sech^2 b (u_a(r)-r_0)}dr-\tanh(cr)]}} .
\end{eqnarray}

\begin{figure}[htb] 
       \begin{minipage}[t]{0.47 \linewidth}
      \fbox{\includegraphics[width=\linewidth]{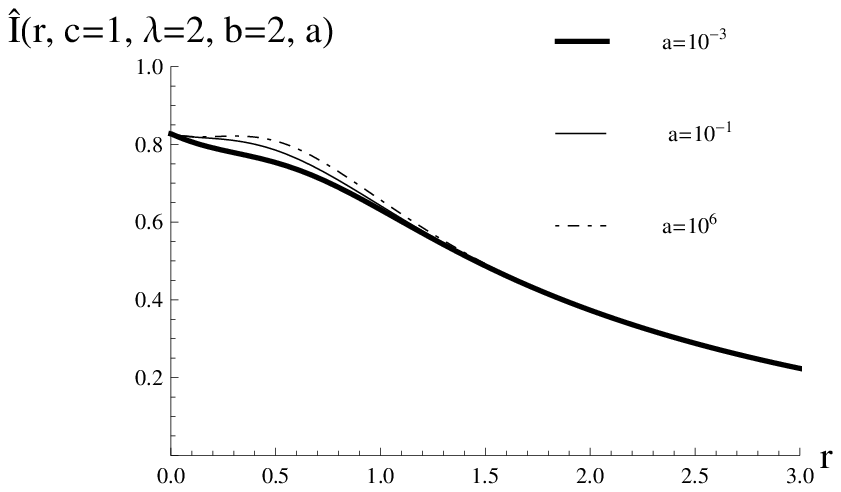}}\\
           \caption{$\hat{I}$ for $a\neq0$ and $r_0=0.2$.}
           \label{normal-figb}\end{minipage}\hfill
       \begin{minipage}[t]{0.48 \linewidth}
           \fbox{\includegraphics[width=\linewidth]{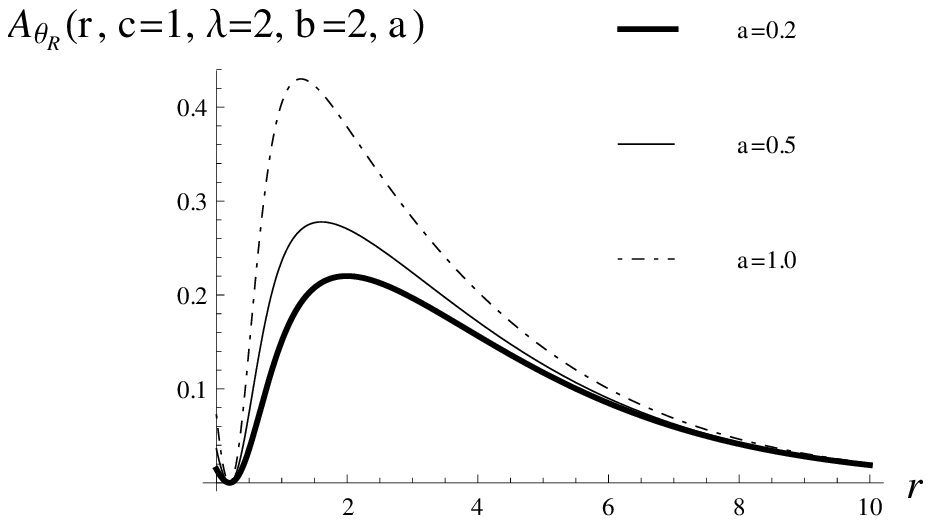}}\\
           \caption{Angular gauge component for $a\neq 0$.}
           \label{atheta-figb-a}
       \end{minipage}
\end{figure}

Once again in order to obtain a normalizable massless mode we choose $\lambda>c$. Moreover, for a smooth solution we need to add the condition
\begin{equation}
 \lambda\int{\sech^2 b(u_a(r)-r_0)}dr > \tanh(cr),
\end{equation}
that constrains the free parameter $\lambda,b$ to $c$. We plotted the radial component $I_{R}$ (\ref{normal2b}) in figure (Figure-\ref{normal-figb}) and the angular gauge component in (Figure-\ref{atheta-figb-a}) for $c=1$, $\lambda=2$ and $r_0=0.2$. Note that as more $b$ more localized are the massless mode and the
angular component.


\subsection{Massive modes}

Now let us turn our attention to the massive modes of eq. (\ref{Dirac-6D-4}). Performing the following change of independent variable $\frac{dz}{dr}=W^{-\frac{1}{2}}(r,c)$, eq. (\ref{Dirac-6D-4}) can be rewritten as
\begin{eqnarray}\label{alphasplit}
\Big[\partial_z + \frac{\dot{W}}{W}+ \frac{\big( \dot{\beta W}\big)}{4\beta W}-iqW^{\frac{1}{2}}A_z\pm\frac{qA_{\theta}}{\sqrt{\beta}}\Big]\Big[\partial_z + \frac{\dot{W}}{W}+ \frac{\dot{\big(\beta W\big)}}{4\beta W}-iqW^{\frac{1}{2}}A_z\mp\frac{qA_{\theta}}{\sqrt{\beta }}
\Big]\alpha_{R,L}(z)=\nonumber\\
=-m^2\alpha_{R,L}(z),
\end{eqnarray}
where the dot ($\cdot$) stands for $\frac{d}{dz}$. A noteworthy feature of eq. (\ref{alphasplit}) is that the only difference between the right-handed and left-handed states lies in the sign of the
gauge coupling.

By means of the change of dependent-variable
\begin{eqnarray}\label{changevaralpha}
\alpha_{R,L}(z)=\exp{\big[-\Big(\frac{5}{4}[cz-\tanh(cz)]-\int_z  iqW^{\frac{1}{2}}(z')A_r(z')dz'\Big)\big]}\tilde{\alpha}_{R,L}(z),
\end{eqnarray}
the eq.(\ref{alphasplit}) turns to be
\begin{eqnarray}
\label{dirac-sch\"{o}rdinger}
\Big(-\partial_z^2 +\Big[\Big(\frac{qA_{\theta}(z)}{\sqrt{\beta(z)}}\Big)^2 \mp \dot{\Big(\frac{qA_{\theta}(z)}{\sqrt{\beta(z)}}\Big)}\Big]\Big)\tilde{\alpha}_{R,L}(z)=m^2\tilde{\alpha}_{R,L}(z).
\end{eqnarray}

Eq. (\ref{dirac-sch\"{o}rdinger}) is a Sch\"{o}rdinger-like equation whose potentials (in the original variable, r) are given by
\begin{eqnarray}
\label{pot1}
V_{R,L}(r)=\Big(\frac{qA_{\theta}(r)}{\sqrt{\beta(r)}}\Big)^2 \mp \sqrt{W(r)}\Big(\frac{qA_{\theta}(r)}{\sqrt{\beta(r)}}\Big)^{'}.
\end{eqnarray}
The expression for the Schr\"{o}dinger-like potential in eq.(\ref{pot1}) is similar to one found in compact $6D$ braneworlds \cite{Parameswaran:2007}. The difference here
is the dependence on the metric components that evolve under the resolution flow.

\subsubsection{Conical case $a=0$}
\begin{figure}[htb] 
       \begin{minipage}[t]{0.48 \linewidth}
           \fbox{\includegraphics[width=\linewidth]{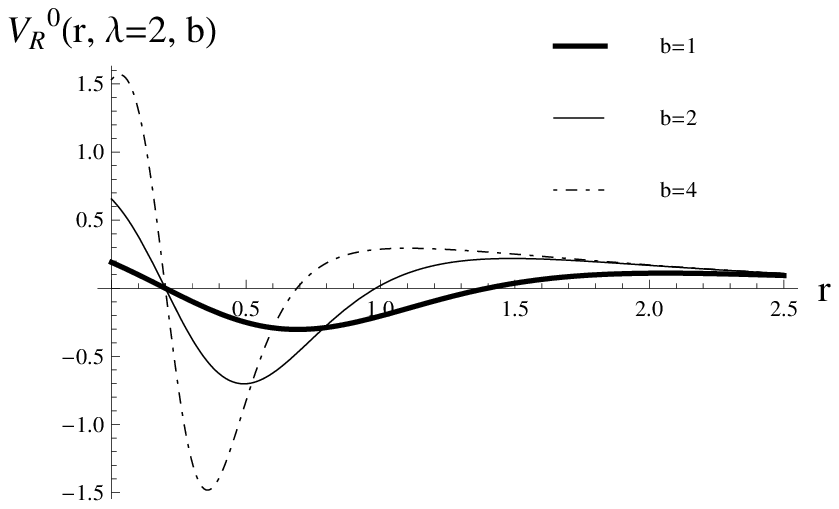}}\\
           \caption{Potential for right-handed fermion and $a=0$.}
           \label{vr-fig}
       \end{minipage}\hfill
       \begin{minipage}[t]{0.48 \linewidth}
           \fbox{\includegraphics[width=\linewidth]{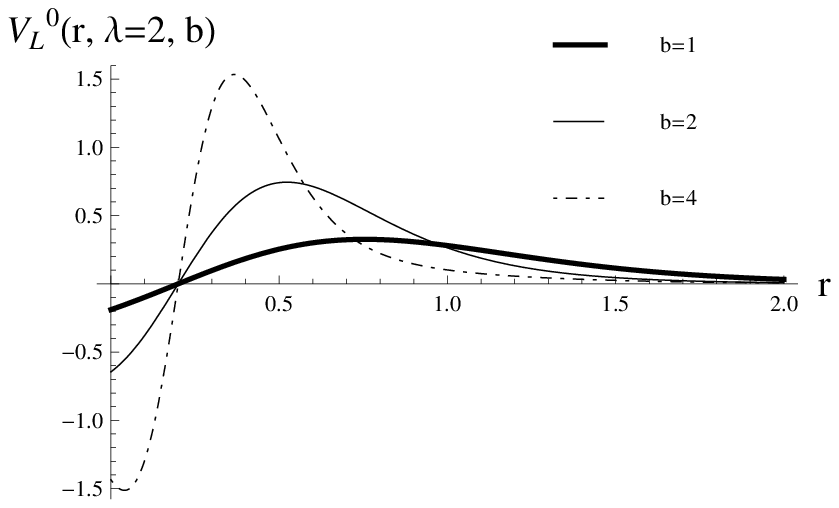}}\\
       \caption{Potential for left-handed fermion and $a=0$.}
           \label{vl-fig}
       \end{minipage}
\end{figure}

For $a=0$, the Sch\"{o}dinger potential takes the form
\begin{eqnarray}\label{pot1}
V_{R,L}^0(r,\lambda,b)=\frac{\lambda^2 }{16} e^{-[c r-\text{tanh}(c r)]}  \text{tanh}b (r-0.2) \Big[\text{tanh}^3 b (r-0.2)+\nonumber \\
\pm \frac{1}{\lambda^2}\Big(8 b \text{sech}^2b (r-0.2)-2 c \text{tanh}^2(c r)\text{tanh}b (r-0.2) \Big)\Big],
\end{eqnarray}
whose graphics were plotted in Figure (\ref{vr-fig}) (for right-handed fermion) and Figure  (\ref{vl-fig}) (for left-handed fermion), both for $c=1$, $\lambda=2$ and $r_0=0.2$.

As usual in braneworld scenarios, only the left-handed potential is attractive at the brane. Moreover, the parameter $b$
controls the depth of the well and the height of the barrier. Furthermore, the potential does not possesses an asymptotic gap, as for 5D thick branes \cite{Casa:2009, Cruz:2011ru, Liu:2009ve, Liu:2009mga}. Therefore, despite the lack
of a massless mode on the brane for $a=0$ (conical behavior), there could be resonant KK modes on the brane.

\subsubsection{Resolved case $a\neq0$}
\begin{figure}[htb] 
        \begin{minipage}[t]{0.48 \linewidth}
           \fbox{\includegraphics[width=\linewidth]{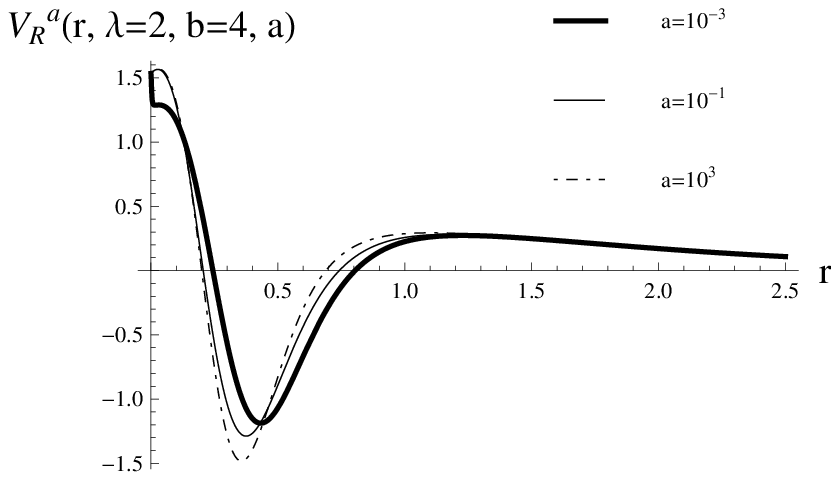}}\\
          \caption{Potential for right-handed fermion and $a\neq0$.}
           \label{vrb2-fig}
       \end{minipage}\hfill
        \begin{minipage}[t]{0.48 \linewidth}
           \fbox{\includegraphics[width=\linewidth]{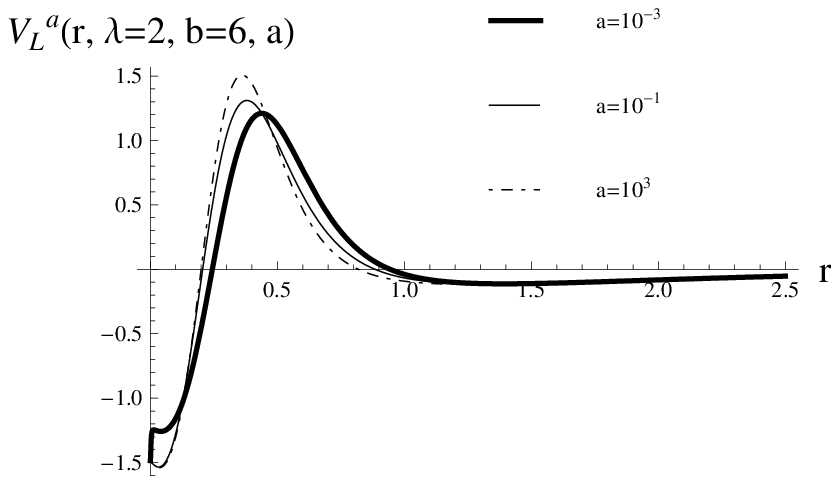}}\\
          \caption{Potential for left-handed fermion and $a\neq0$.}
           \label{vlb2-fig}
       \end{minipage}
\end{figure}

For $a\neq 0$, the potential turns to be
\begin{eqnarray}\label{pot1b}
V_{R,L}^a(r,\lambda,b)=\frac{\lambda^2}{16}e^{-[c r-\tanh(c r)]}\tanh b(u_a(r)-0.2)\Big[\tanh^{3}b(u_a(r)-0.2)\mp\frac{16a^2b}{\lambda(r^2+6a^2)}\times\nonumber\\
\times \Big(\sqrt{\frac{(r^2+6a^2)(r^2+9a^2)}{54a^2}}\text{sech}^2b(u_a(r)-0.2)-2c\tanh^2(cr)\tanh b(u_a(r)-0.2)\Big)\Big],
\end{eqnarray}
that are represented in the Figure  (\ref{vrb2-fig}) (l.h. fermion) and in the Figure (\ref{vlb2-fig}) (r.h. fermion), both for $c=1$, $\lambda=2$ and $r_0=0.2$.

Again, the potential has an usual volcano shape for only the left-handed fermions. Note that as higher the value of $a$ higher the barrier. Furthermore, for $a\approx 0$ the potential has an abrupt change near the origin.

\section{Conclusions and perspectives}
\label{Conclusions and perspectives}

In this work we have analyzed the effects of a resolution flow upon the geometry and fermion field defined on a warped braneworld scenario built from a $3$-brane and a $2$-cycle of the resolved conifold.

Firstly, we have shown that this ansatz smoothes out the well-known string-like geometry near the brane and retrieves the string-like properties asymptotically.

The next step was to study how a massless fermion behaves in this geometrical flow. For a minimally coupled $l=0$ states with a background gauge vector field, we turned out that the massless mode is ill-defined on the brane for $a=0$, due the conical behavior. The resolution parameter solves this problem by allowing a chiral zero-mode whose value on the brane depends on $a$.

On the other hand, we showed that for the massive modes, the Schr\"{o}dinger-like potential has the usual volcano-shape for the left-handed fermion, for both $a=0$ and $a\neq0$. For $a\approx 0$ the potential has an abrupt change near the origin.

For future works we intend to study the solutions for $l\neq0$. Moreover, another important issue to be addressed are the effects of the resolution parameter has on the resonant modes with or without a background vector field.

\section{Acknowledgments}

Financial support from the Brazilian agencies Conselho Nacional de Desenvolvimento Cient\'{\i}fico e
Tecnol\'{o}gico (CNPq) and  Coordena\c{c}\~{a}o de Aperfei\c{c}oamento de Pessoal de N\'{i}vel Superior (CAPES) is gratefully acknowledged.


\begin{thebibliography}{99}
\bibliographystyle{unsrt}
\bibliographystyle{alpha}

\bibitem{Randall:1999ee}
  L.~Randall, and R.~Sundrum,
  Phys.\ Rev.\ Lett.\  {\bf 83}, 3370 (1999)
  [hep-ph/9905221].

\bibitem{Randall:1999vf}
  L.~Randall and R.~Sundrum,
  Phys.\ Rev.\ Lett.\  {\bf 83}, 4690 (1999)
  [arXiv:hep-th/9906064].


\bibitem{Kehagias:2000au}
  A.~Kehagias and K.~Tamvakis,
  Phys.\ Lett.\ B {\bf 504}, 38 (2001)
  [hep-th/0010112].

\bibitem{Olasagasti:2000gx}
  I.~Olasagasti and A.~Vilenkin,
  Phys.\ Rev.\ D {\bf 62}, 044014 (2000)
  [hep-th/0003300].

\bibitem{Liu:2007gk}
  Y.~-X.~Liu, L.~Zhao and Y.~-S.~Duan,
  JHEP {\bf 0704}, 097 (2007)
  [hep-th/0701010].

\bibitem{Oda:2000zc}
  I.~Oda,
  Phys.\ Lett.\  B {\bf 496}, 113 (2000)
  [arXiv:hep-th/0006203].

\bibitem{Gherghetta:2000qi}
  T.~Gherghetta and M.~E.~Shaposhnikov,
  Phys.\ Rev.\ Lett.\  {\bf 85}, 240 (2000)
  [arXiv:hep-th/0004014].


\bibitem{Cohen:1999ia}
  A.~G.~Cohen and D.~B.~Kaplan,
  Phys.\ Lett.\  B {\bf 470}, 52 (1999)
  [arXiv:hep-th/9910132].

\bibitem{Gregory:1999gv}
  R.~Gregory,
  Phys.\ Rev.\ Lett.\  {\bf 84}, 2564 (2000)
  [arXiv:hep-th/9911015].



\bibitem{Giovannini:2001hh}
  M.~Giovannini, H.~Meyer and M.~E.~Shaposhnikov,
  Nucl.\ Phys.\ B {\bf 619}, 615 (2001)
  [hep-th/0104118].




\bibitem{Tinyakov:2001jt}
  P.~Tinyakov and K.~Zuleta,
  Phys.\ Rev.\ D {\bf 64}, 025022 (2001)
  [hep-th/0103062].




\bibitem{Ponton:2000gi}
  E.~Ponton and E.~Poppitz,
  JHEP {\bf 0102}, 042 (2001)
  [arXiv:hep-th/0012033].





\bibitem{Bostock:2003cv}
  P.~Bostock, R.~Gregory, I.~Navarro and J.~Santiago,
  Phys.\ Rev.\ Lett.\  {\bf 92}, 221601 (2004)
  [hep-th/0311074].

\bibitem{Candelas:1989js}
  P.~Candelas and X.~C.~de la Ossa,
  Nucl.\ Phys.\  B {\bf 342}, 246 (1990).

\bibitem{Greene:1995hu}
  B.~R.~Greene, D.~R.~Morrison and A.~Strominger,
  Nucl.\ Phys.\  B {\bf 451}, 109 (1995)
  [arXiv:hep-th/9504145].

\bibitem{p} J.Polchinski, String theory and beyond, vol.2, Cambrigde University press.

\bibitem{Pando Zayas:2000sq}
  L.~A.~Pando Zayas and A.~A.~Tseytlin,
  JHEP {\bf 0011}, 028 (2000)
  [arXiv:hep-th/0010088].

\bibitem{Cvetic:2000mh}
  M.~Cvetic, H.~Lu and C.~N.~Pope,
  Nucl.\ Phys.\  B {\bf 600}, 103 (2001)
  [arXiv:hep-th/0011023].


\bibitem{Minasian:1999tt}
  R.~Minasian and D.~Tsimpis,
  Nucl.\ Phys.\ B {\bf 572}, 499 (2000)
  [hep-th/9911042].
\bibitem{Klebanov:2007us}
  I.~R.~Klebanov and A.~Murugan,
  JHEP {\bf 0703}, 042 (2007)
  [hep-th/0701064].



\bibitem{VazquezPoritz:2001zt}
  J.~F.~Vazquez-Poritz,
  JHEP {\bf 0209}, 001 (2002)
  [arXiv:hep-th/0111229].


\bibitem{Klebanov:1999tb}
  I.~R.~Klebanov and E.~Witten,
  Nucl.\ Phys.\ B {\bf 556}, 89 (1999)
  [hep-th/9905104].


\bibitem{Klebanov:2000hb}
  I.~R.~Klebanov and M.~J.~Strassler,
  JHEP {\bf 0008}, 052 (2000)
  [arXiv:hep-th/0007191].

\bibitem{Klebanov:2000nc}
  I.~R.~Klebanov and A.~A.~Tseytlin,
  Nucl.\ Phys.\ B {\bf 578}, 123 (2000)
  [hep-th/0002159].



\bibitem{Silva:2011yk}
  J.~E.~G.~Silva, C.~A.~S.~Almeida,
  Phys.\ Rev.\ D {\bf 84}, 085027 (2011)
  [arXiv:1110.1597 [hep-th]].

\bibitem{Silva:2012yj}
  J.~E.~G.~Silva, V.~Santos, C.~A.~S.~Almeida and ,
  Class.\ Quant.\ Grav.\  {\bf 30}, 025005 (2013)
  [arXiv:1208.2364 [hep-th]].

\bibitem{WILLIAMS:2012AU}
  M.~Williams, C.~P.~Burgess, L.~Van Nierop and A.~Salvio,
  JHEP {\bf 1301}, 102 (2013)
  [arXiv:1210.3753 [hep-th]].


\bibitem{Costa:2013}
  F.~W.~V.~Costa, J.~E.~G.~Silva and C.~A.~S.~Almeida,
  Phys.\  Rev.\  D {\bf 87}, 125010 (2013)
  [arXiv:1304.7825 [hep-th]].





\bibitem{Parameswaran:2007}
  S.~L.~Parameswaran, S.~ Randjbar-Daemi, A.~ Salvio,
  Nucl.\ Phys.\  B {\bf 767}, 54-81 (2007)
  [arXiv:hep-th/0608074 ].

\bibitem{Neronov:2001}
 A.~ Neronov
   Phys.\ Rev.\ D {\bf 65}, 044004 (2001)
  [arXiv:0106092v1 [gr-qc]].



   \bibitem{Merab:2007}
  M.~ Gogberashvili, P.~ Midodashvili, D.~ Singleton,
   JHEP {\bf 0708}, 033 (2007)
  [arXiv:0706.0676v2 [hep-th]].

\bibitem{Aguilar:2006}
 S.~ Aguilar, D.~ Singleton
   Phys.\ Rev.\ D {\bf 73}, 085007 (2006)
  [arXiv:0602218v3 [hep-th]].

\bibitem{Casa:2009}
C.~ A.~ S.~ Almeida, M.~ M.~ Ferreira~ Jr., A.~ R.~ Gomes, R.~ Casana
   Phys.\ Rev.\ D {\bf 79}, 125022 (2009)
  [arXiv:0901.3543v2 [hep-th]].


\bibitem{Cruz:2011ru}
  W.~T.~Cruz, A.~R.~Gomes and C.~A.~S.~Almeida,
  Eur.\ Phys.\ J.\ C {\bf 71}, 1790 (2011)
  [arXiv:1110.4651 [hep-th]].

\bibitem{Liu:2009ve}
  Y.~-X.~Liu, J.~Yang, Z.~-H.~Zhao, C.~-EFu and Y.~-S.~Duan,
  Phys.\ Rev.\ D {\bf 80}, 065019 (2009)
  [arXiv:0904.1785 [hep-th]].

\bibitem{Liu:2009mga}
  Y.~-X.~Liu, H.~-T.~Li, Z.~-H.~Zhao, J.~-X.~Li and J.~-R.~Ren,
  JHEP {\bf 0910}, 091 (2009)
  [arXiv:0909.2312 [hep-th]].

\end{thebibliography}
\end{document}